\pdfoutput=1
\documentclass[conference]{IEEEtran}
\IEEEoverridecommandlockouts

\usepackage{amsmath,amssymb,amsfonts}
\usepackage{algorithmic}
\usepackage{algorithm}
\usepackage{graphicx}
\usepackage{textcomp}
\usepackage{xcolor}
\usepackage[nocompress]{cite}
\usepackage{bm}
\usepackage{subfig}
\usepackage{multirow}
\newcommand{\vect}[1]{\boldsymbol{#1}}
\newcommand{\mat}[1]{\mathbf{#1}}
\newcommand{\set}[1]{\mathcal{#1}}

\newcommand{\complex}{\mathbb{C}}

\newcommand{\diag}{\mathrm{diag}}

\newcommand{\trace}{\mathrm{tr}}
\DeclareMathOperator*{\argmax}{arg\,max}

\usepackage{fancyhdr}
\fancypagestyle{firststyle}{\fancyhf{}
		\fancyhead[L]{M. Ying, G. Qian, X. Wang, X. Liu, I. S. Gupte, D. Shakya, and T. S. Rappaport, ``PRIMO: GPU-Accelerated Joint Beamforming and RIS Placement Optimization for Outdoor-to-Indoor Coverage at FR1(C) and FR3," in \textit{Asilomar Conference on Signals, Systems, and Computers}, Pacific Grove, CA, USA, Oct. 25--28, 2026.}
	}

\begin{document}
\bstctlcite{BSTcontrol}

\title{PRIMO: GPU-Accelerated Joint Beamforming and RIS Placement Optimization for Outdoor-to-Indoor Coverage at FR1(C) and FR3}

\author{\IEEEauthorblockN{Mingjun Ying$^{1,*}$, Guanyue Qian$^{2,*}$, Xinquan Wang, Xingchen Liu, Ishaan S. Gupte, \\Dipankar Shakya and Theodore S. Rappaport$^{3}$\thanks{$^{*}$Mingjun Ying and Guanyue Qian contributed equally to this work.}}
\IEEEauthorblockA{NYU WIRELESS, New York University, Brooklyn, NY, USA \\
$^{1}$yingmingjun@nyu.edu, $^{2}$gq2032@nyu.edu, $^{3}$tsr@nyu.edu}}

\maketitle
\thispagestyle{firststyle}

\begin{abstract}
Upper mid-band spectrum is promising for 5G and 6G cellular networks, yet concrete and low-emissivity glass impose over 20 dB of building penetration loss above 2 GHz, degrading outdoor-to-indoor (O2I) coverage. A reconfigurable intelligent surface (RIS) on a neighboring building bypasses this blockage by focusing energy onto indoor users. This paper introduces PRIMO (Parallel RIS Min-Rate Optimization), a training-free joint optimizer of base-station precoding, RIS phase shifts, and RIS placement, validated by calibrated ray tracing in a MetroTech Commons digital twin (Brooklyn, NY) at 6.75 and 16.95 GHz. At equal physical aperture, half-wavelength spacing packs 6.25 times as many elements per array at FR3 as at FR1(C), so FR1(C) leads at low indoor-user counts while FR3 dominates at higher counts, where the additional spatial degrees of freedom suppress inter-user interference. Above a 1 bps/Hz service floor, PRIMO supports 92\% more indoor users at FR3 than at FR1(C), and raises the minimum user rate by up to 90\% over a fully converged alternating-optimization baseline while running about 100 times faster.
\end{abstract}


\section{Introduction}
\label{sec:intro}

More than 80\% of wireless data traffic originates or terminates indoors~\cite{rappaport2019_above100ghz}, yet cellular operators cannot guarantee indoor coverage. Energy-efficient building materials, including low-emissivity glass and reinforced concrete, impose building penetration loss (BPL) exceeding 20~dB above 2~GHz in modern construction~\cite{itu_p2040, rappaport2024_fr1c}, and distributed antenna systems (DAS) that could restore indoor coverage are costly to deploy.

The World Radiocommunication Conference 2023 identified the 6.425--7.125~GHz band (FR1(C)) for International Mobile Telecommunications globally, while the 3rd Generation Partnership Project (3GPP) is studying FR3 (7.125--24.25~GHz) for future cellular use~\cite{3gpp_tr38913, 3gpp_tr38901}; recent measurements characterize indoor~\cite{rappaport2024_fr1c}, outdoor urban~\cite{shakya2025_umi_icc}, and factory~\cite{ying2025_inf_icc} channels at both ranges. A reconfigurable intelligent surface (RIS) provides a cost-effective passive alternative to DAS: each element applies a controllable phase shift, redirecting signals through lower-loss materials such as glass rather than concrete~\cite{wu2021_ris_survey, di_renzo2020_ris}, a glass-portal strategy first articulated for mmW O2I in~\cite{rappaport2020_interface_points} and consistent with measured cascaded path-loss models~\cite{tang2021_ris_pathloss}.

Prior RIS beamforming research uses alternating optimization with element-wise phase update~\cite{wu2019_ris_bf}, multi-cell multiple-input multiple-output (MIMO) joint precoding~\cite{pan2020_ris_multicell}, weighted sum-rate maximization~\cite{guo2020_ris_weighted}, manifold optimization~\cite{yu2020_manifold_ris,gmml}, gradient-based path optimization~\cite{gbml}, max-min SINR gradient ascent~\cite{nadeem2020_ris_ojcoms}, or semidefinite relaxation (SDR)~\cite{luo2010_sdr}, all assuming a fixed RIS position. Deployment studies span analytical placement~\cite{zeng2021_ris_placement}, whose simplified path-loss models miss material-dependent BPL, max-min placement over random users~\cite{rajasekaran2026_ris_placement}, and calibrated city-scale ray-tracing deployment~\cite{beyraghi2025_ris_deploy} jointly optimizing RIS placement, orientation, and BS beamforming. Our work complements~\cite{beyraghi2025_ris_deploy} with a single-RIS multi-user outdoor-to-indoor (O2I) downlink under calibrated ITU-R P.2040 material penetration losses and a worst-user fairness objective at upper mid-band, where O2I penetration loss increases by approximately 8~dB from FR1(C) to FR3~\cite{rappaport2024_fr1c}, where RIS gains are scenario-dependent~\cite{kara2026_ris_fr3}, making coverage guarantees more difficult precisely where spectrum is newly available. Calibrated ray-tracing (RT) tools such as NYURay~\cite{kanhere2025_nyuray} combined with site-specific 3D digital twins~\cite{ying2026_horama} model O2I propagation with high fidelity, yet joint optimization of precoding, RIS phases, and deployment position at upper mid-band has not been studied; this paper addresses that gap at 6.75 and 16.95~GHz.

The main contributions are:
\begin{itemize}
    \item We propose PRIMO (Parallel RIS Min-Rate Optimization), a training-free joint optimizer of base-station precoding, RIS phase shifts, and RIS deployment position, which extends gradient-based max-min designs~\cite{nadeem2020_ris_ojcoms} with log-sum-exp smoothing of the minimum user rate, GPU-batched parallel restarts with progressive pruning, and per-step recomputation of the regularized zero-forcing precoder. PRIMO improves the minimum user rate by up to a factor of $1.90$ ($2.26$ vs.\ $1.19$~bps/Hz, mean of 5 trials) over a fully converged alternating-optimization baseline at FR3 with 16 indoor users, while converging about 100 times faster on a single GPU ($57$ vs.\ $6133$~s at FR3, $K\!=\!8$).

    \item We present a calibrated dual-band ray-tracing study of RIS-aided O2I coverage in a MetroTech Commons digital twin with both BS and RIS held at the same physical aperture across bands, so half-wavelength packing yields about $6.25\times$ more elements at FR3 (BS $M\!=\!256\!\to\!1600$, RIS $N\!=\!1024\!\to\!6400$). Sweeping the load from 4 to 20 indoor users shows the dual-band advantage is load-dependent, with FR3 delivering a $4.70\times$ minimum user rate over FR1(C) at 12 indoor users ($6.49$ vs.\ $1.38$~bps/Hz), where the larger element count suppresses inter-user interference (IUI).

    \item We sweep the RIS position along a 40-meter facade track and identify deployment geometry as a first-order design variable, decided once per site over the served indoor area rather than per scheduling instant. Adjacent-facade placement with outdoor LoS to the BS minimizes cascaded path loss, and RIS phase optimization alone cannot recover from a poor position.
\end{itemize}

\section{System Model}
\label{sec:system_model}

\begin{figure}[t]
    \centering
\includegraphics[width=0.72\columnwidth]{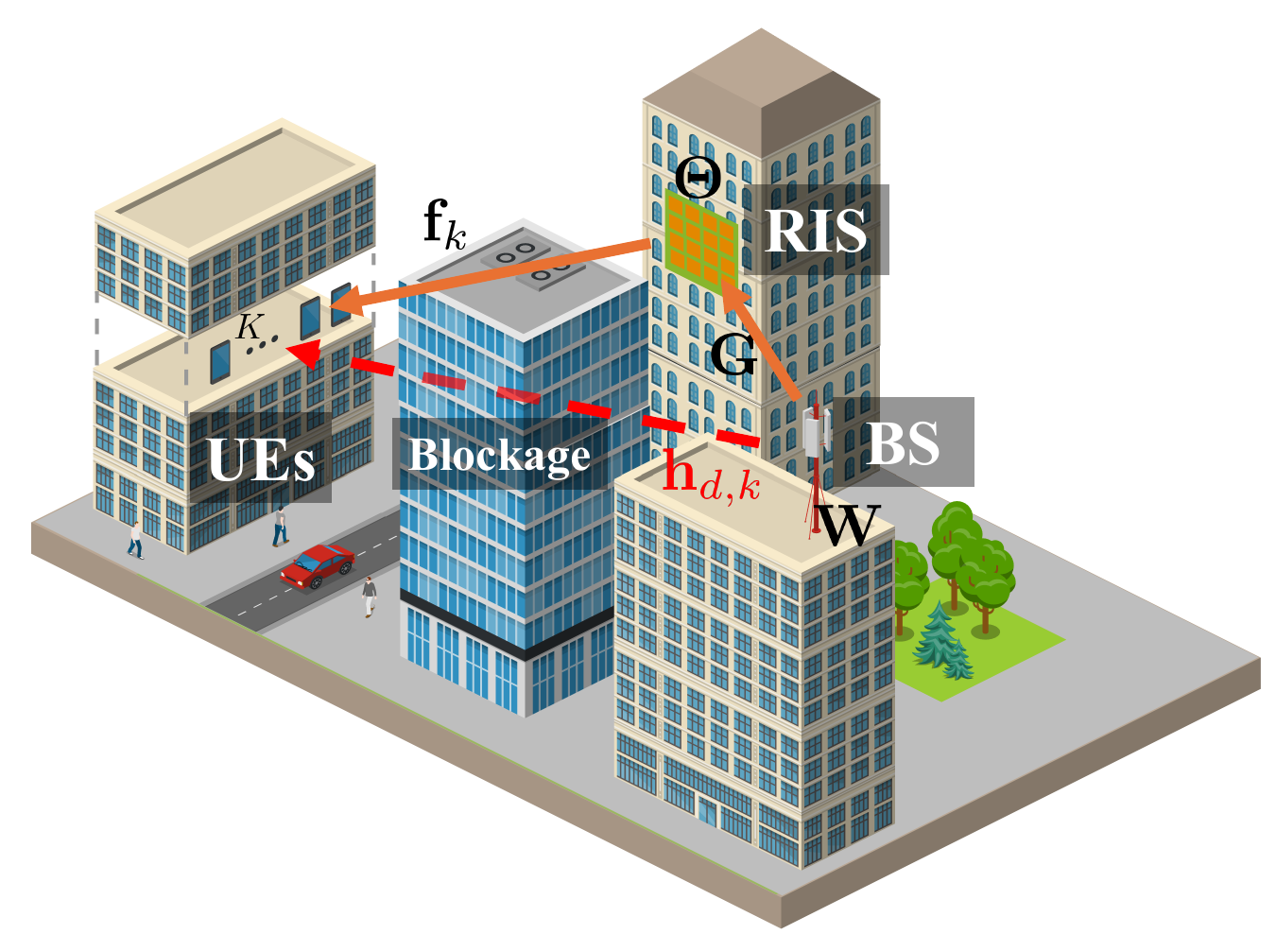}
    \caption{O2I MU-MISO system model. The direct path $\mathbf{h}_{d,k}$ suffers concrete BPL~\cite{itu_p2040}; the RIS-assisted path bypasses the blockage via the BS-to-RIS LoS channel $\mathbf{G}$ and the RIS-to-UE channel $\mathbf{f}_k$ through low-loss glass.}
    \label{fig:system_model}
    \vspace{-10pt}
\end{figure}

Consider a downlink multi-user multiple-input single-output (MU-MISO) system where a BS with $M$ antennas serves $K$ single-antenna indoor user equipments (UEs), as illustrated in Fig.~\ref{fig:system_model}. A concrete building between the BS and the target building blocks the direct BS-to-UE path, causing severe BPL. A passive RIS with $N$ elements is deployed outdoors with line-of-sight (LoS) to the BS, redirecting the BS signal toward the target building through a glass facade with substantially lower BPL. The RIS position $d \in [0, L]$ is a deployment-time design variable selected once for the served indoor area along an $L\!=\!40$~m track parallel to the glass facade, discretized into $N_d\!=\!10$ candidates $\{d_i\}_{i=1}^{N_d}$.

The effective channel from the BS to UE~$k$ combines a direct O2I path and a RIS-assisted path:
\begin{equation}
    \vect{h}_k^H(d, \vect{\theta}) = \vect{h}_{d,k}^H + \big(\vect{f}_k(d)\big)^{\!H} \mat{\Theta} \mat{G}(d),
    \label{eq:h_eff}
\end{equation}
where $\vect{h}_{d,k} \in \complex^M$ denotes the direct BS-to-UE channel attenuated by concrete BPL, $\mat{G}(d) \in \complex^{N \times M}$ denotes the BS-to-RIS channel (outdoor LoS), $\vect{f}_k(d) \in \complex^N$ denotes the RIS-to-UE~$k$ channel penetrating the glass facade, and $\mat{\Theta} = \diag(e^{j\theta_1}, \ldots, e^{j\theta_N})$ denotes the RIS phase-shift matrix.

The BS serves all $K$ UEs with precoding matrix $\mat{W} = [\vect{w}_1, \ldots, \vect{w}_K] \in \complex^{M \times K}$ subject to total power $\trace(\mat{W}^H\mat{W}) \leq P$. The received signal at UE~$k$ is $y_k = \vect{h}_k^H \sum_{j=1}^{K} \vect{w}_j s_j + n_k$, where $s_j \sim \set{CN}(0, 1)$ is the data symbol for UE~$j$ and $n_k \sim \set{CN}(0, \sigma^2)$ is the receiver noise, with thermal noise power $\sigma^2\!=\!k_B T_0\,W\,F$ for bandwidth $W$ and noise figure $F$ (Table~\ref{tab:params}). The signal-to-interference-plus-noise ratio (SINR) of UE~$k$ is
\begin{equation}
    \gamma_k = \frac{|\vect{h}_k^H \vect{w}_k|^2}{\sum_{j \neq k} |\vect{h}_k^H \vect{w}_j|^2 + \sigma^2},
    \label{eq:sinr}
\end{equation}
and the achievable rate is $R_k = \log_2(1 + \gamma_k)$.

\section{Problem Formulation and Proposed Solutions}
\label{sec:solutions}

\subsection{Max-Min Rate Problem}

In O2I scenarios, indoor UEs experience highly heterogeneous channels due to position-dependent BPL: UEs behind glass receive substantially stronger signals than UEs behind concrete, so a sum-rate objective would leave deep-fade UEs unserved. We therefore maximize the minimum user rate over BS precoding $\mat{W}$, RIS phase shifts $\vect{\theta}$, and RIS position $d$:
\begin{subequations}
\label{eq:P1}
\begin{align}
    (\text{P1}): \quad \max_{\mat{W}, \vect{\theta}, d} \quad & \min_{k \in \{1, \ldots, K\}} \; R_k(\mat{W}, \vect{\theta}, d) \label{eq:P1_obj} \\
    \text{s.t.} \quad & \trace(\mat{W}^H\mat{W}) \leq P, \label{eq:P1_power} \\
    & |e^{j\theta_n}| = 1, \;\; n = 1, \ldots, N, \label{eq:P1_unit} \\
    & 0 \leq d \leq L. \label{eq:P1_pos}
\end{align}
\end{subequations}

Problem (P1) is non-convex due to (i) the unit-modulus constraints~\eqref{eq:P1_unit} forming the manifold $\set{S}^N$, (ii) the coupling among $\mat{W}$, $\vect{\theta}$, and $d$ through~\eqref{eq:h_eff}, and (iii) the non-smooth inner $\min_k$ operator; no closed-form joint solution exists.

The standard alternating optimization (AO) approach~\cite{wu2019_ris_bf} solves (P1) by cycling through three blocks per iteration: the regularized zero-forcing (RZF) precoder $\mat{W}$ for fixed phases and position, element-wise (EW) phase updates for fixed $\mat{W}$ and $d$, and selection of the best $d$ among $N_d$ candidates. For O2I, $d$ is a genuine third block, since it sets which material (concrete or glass) the RIS-to-UE path crosses.

\subsection{Proposed PRIMO: Parallel RIS Min-Rate Optimization}
\label{sec:primo}

AO-EW~\cite{wu2019_ris_bf} has three structural limitations: (i)~the RZF precoder uses equal power allocation (PA), which is suboptimal for the max-min objective, (ii)~the element-wise phase update sweeps $N$ phases sequentially at $\set{O}(N)$ per pass, and (iii)~the block-coordinate structure fixes $\mat{W}$ during all $N$ phase updates, causing slow alternating descent between complementary sub-problems. PRIMO addresses all three limitations as a GPU-parallel batch optimizer that eliminates the block-coordinate structure entirely.

For fixed $\vect{\theta}$ and $d$, the effective channels $\vect{h}_k$ in~\eqref{eq:h_eff} are known and, following~\cite{wu2019_ris_bf, guo2020_ris_weighted}, the RZF precoder is:
\begin{equation}
    \tilde{\mat{W}} = \mat{H}^H(\mat{H}\mat{H}^H + \alpha \mat{I}_K)^{-1}, \quad \alpha = K\sigma^2/P,
    \label{eq:rzf}
\end{equation}
where $\mat{H} = [\vect{h}_1, \ldots, \vect{h}_K]^H$, and $\mat{W} = \sqrt{P}\,\tilde{\mat{W}}/\|\tilde{\mat{W}}\|_F$. Unlike AO-EW, which recomputes $\mat{W}$ once per outer iteration, PRIMO refreshes $\mat{W}$ at every gradient step, keeping the precoder tightly coupled to the current phases; this closed-form per-step RZF with first-order phase updates avoids both an inner convex solve and implicit differentiation, making GPU batching over hundreds of restarts tractable.

The next design element is a smooth, batched phase update. The non-smooth $\min_k$ operator in~\eqref{eq:P1_obj} prevents direct gradient computation. PRIMO therefore replaces the $\min_k$ operator with the following smooth log-sum-exp surrogate $\set{L}(\vect{\theta})$, with smoothing parameter $\alpha_{\mathrm{sm}}\!>\!0$,
\begin{equation}
    \set{L}(\vect{\theta}) = -\frac{1}{\alpha_{\mathrm{sm}}}\log\!\left(\sum_{k=1}^{K} e^{-\alpha_{\mathrm{sm}} R_k(\vect{\theta})}\right),
    \label{eq:smooth_min}
\end{equation}
a lower bound satisfying $\min_k R_k - \log K/\alpha_{\mathrm{sm}} \leq \set{L} \leq \min_k R_k$; we set $\alpha_{\mathrm{sm}}\!=\!15$, a gap of at most $\log(K)/15 \approx 0.14$~bps/Hz at $K\!=\!8$. PRIMO updates all $N$ phases simultaneously, differentiating~\eqref{eq:smooth_min} with respect to the unit-modulus variables $\vect{\phi}\!=\![e^{j\theta_1},\ldots,e^{j\theta_N}]^T$ via \emph{Wirtinger calculus}~\cite{kreutz2009_wirtinger} as
\begin{equation}
    \nabla_{\vect{\phi}^*}\!\set{L} = \sum_{k=1}^{K} \frac{p_k}{\ln 2\,(1\!+\!\gamma_k) I_k} \left[ s_{k,k}\, \vect{a}_{k,k} - \gamma_k \!\sum_{j \neq k} s_{k,j}\, \vect{a}_{k,j} \right],
    \label{eq:primo_grad}
\end{equation}
where $p_k = e^{-\alpha_{\mathrm{sm}} R_k}/\sum_j e^{-\alpha_{\mathrm{sm}} R_j}$ are softmax weights concentrating the gradient on the worst UE, $s_{k,j} = \vect{h}_k^H \vect{w}_j$, $\vect{a}_{k,j} = \vect{f}_k^* \odot (\mat{G}\vect{w}_j)$, $I_k = \sum_{j \neq k} |s_{k,j}|^2 + \sigma^2$, and $\gamma_k$ is the SINR. The interference term $-\gamma_k \sum_{j \neq k} s_{k,j} \vect{a}_{k,j}$ steers $\vect{\theta}$ away from solutions that boost one UE at the cost of others. PRIMO uses Adam~\cite{kingma2015_adam} with cosine-annealed step size $\eta_0\!=\!0.05$ over $T\!=\!T_1\!+\!T_2$ steps in two progressive-pruning stages: $B\!=\!500$ parallel restarts run $T_1\!=\!30$ screening steps, then only the top $\rho\!=\!10\%$ are retained for $T_2\!\approx\!270$ refinement steps, reducing computation by $\sim$5$\times$ versus full-batch. At each step, $\mat{W}$ is treated as $\vect{\theta}$-independent during backpropagation and refreshed via~\eqref{eq:rzf} before the next step, rather than implicitly differentiated. The gradient~\eqref{eq:primo_grad} costs $\set{O}(NK^2M)$ per step.

After phase convergence at each candidate position, PRIMO selects the position $d^{\star}$ achieving the highest $\min_k R_k$ and applies max-min SINR power allocation~\cite{schubert2004_sinr} to equalize SINR across UEs, yielding $\vect{w}_k = \sqrt{\pi_k^{\star}} \hat{\vect{w}}_k$ with equal-SINR power coefficients $\{\pi_k^{\star}\}$.

For slowly-varying O2I channels such as pedestrian indoor UEs, PRIMO tracks channel variation by initializing from the previous slot's $\vect{\theta}^{\star}$ with $B\!=\!1$ and $T\!=\!10$ (warm-start mode), reducing latency to $\approx$2~ms on a single graphics processing unit (GPU), well within the 10--80~ms beam management periodicity of 5G New Radio. Algorithm~\ref{alg:primo} summarizes PRIMO.

\renewcommand{\algorithmicrequire}{\textbf{Input:}}
\renewcommand{\algorithmicensure}{\textbf{Output:}}
\begin{algorithm}[t]
\caption{PRIMO: Parallel RIS Min-Rate Optimization}
\label{alg:primo}
\small
\begin{algorithmic}[1]
\REQUIRE Channels $\{\mat{G}(d_i),\vect{f}_k(d_i),\vect{h}_{d,k}\}_{i\in[N_d],\,k\in[K]}$; restart count $B$, screening steps $T_1$, refinement steps $T_2$, pruning fraction $\rho\!\in\!(0,1]$, smoothing parameter $\alpha_{\mathrm{sm}}\!>\!0$, base learning rate $\eta_0$
\ENSURE $d^{\star}$, $\mat{W}^{\star}$, $\vect{\theta}^{\star}$
\FOR{$i = 1$ \textbf{to} $N_d$}
    \STATE Sample $\{\vect{\theta}^{(b)}_0\}_{b=1}^{B}\!\sim\!\set{U}[0,2\pi)^N$, reset Adam moments
    \FOR{$t = 1$ \textbf{to} $T_1$ \textit{(stage 1, screen $B$ restarts)}}
        \STATE Form $\{\vect{h}_k^{(b)}\}$ via~\eqref{eq:h_eff}, update $\mat{W}^{(b)}$ via~\eqref{eq:rzf}
        \STATE Compute $\vect{g}^{(b)}\!=\!\nabla_{\vect{\phi}^{*}}\set{L}^{(b)}$ from~\eqref{eq:primo_grad}
        \STATE $\vect{\theta}^{(b)}_t\!\leftarrow\!\textsc{Adam}(\vect{\theta}^{(b)}_{t-1},\vect{g}^{(b)},\eta_t)$, with cosine LR $\eta_t$
    \ENDFOR
    \STATE $\set{B}'\!\leftarrow\!\textsc{TopK}\big(\{\min_k R_k(\vect{\theta}^{(b)}_{T_1})\}_{b=1}^{B},\lceil\rho B\rceil\big)$
    \FOR{$t = T_1\!+\!1$ \textbf{to} $T_1\!+\!T_2$ \textit{(stage 2, refine)}}
        \STATE Repeat lines~4--6 restricted to $b\!\in\!\set{B}'$
    \ENDFOR
    \STATE $b^{\star}_i\!\leftarrow\!\argmax_{b\in\set{B}'}\min_k R_k(\vect{\theta}^{(b)}_{T_1+T_2})$
    \STATE Apply max-min SINR power allocation~\cite{schubert2004_sinr}
    \STATE $R_{\min}(d_i)\!\leftarrow\!\min_k R_k(\vect{\theta}^{(b^{\star}_i)}_{T_1+T_2})$
\ENDFOR
\STATE $i^{\star}\!\leftarrow\!\argmax_{i\in[N_d]} R_{\min}(d_i)$
\RETURN $d^{\star}\!=\!d_{i^{\star}}$, $\mat{W}^{\star}\!=\!\mat{W}(d_{i^{\star}})$, $\vect{\theta}^{\star}\!=\!\vect{\theta}^{(b^{\star}_{i^{\star}})}(d_{i^{\star}})$
\end{algorithmic}
\end{algorithm}


PRIMO is compared against four classical block-coordinate methods, all using the three-block AO structure (precoding $\to$ phase $\to$ position) with equal-power RZF and 30 random restarts: AO-EW~\cite{wu2019_ris_bf}, which updates the $N$ phases one element at a time under fairness weights $w_k = 1/R_k$; BCD-SDR~\cite{luo2010_sdr}, which relaxes the unit-modulus constraints via semidefinite relaxation with Gaussian randomization; WMMSE+AO~\cite{shi2011_wmmse}, which reformulates weighted sum-rate maximization as a weighted MSE problem with closed-form block updates; and Manifold-RCG~\cite{yu2020_manifold_ris}, which optimizes the phase vector on $\set{S}^N$ by Riemannian conjugate gradient. As a learning-based baseline we adopt the gradient-based liquid neural network (GLNN)~\cite{glnn_wcl2024}, the published learning approach closest to our setting, mapping CSI to $(\hat{d}, \hat{\vect{\theta}})$ end-to-end with the smooth min-rate loss~\eqref{eq:smooth_min} and $20\times$ phase augmentation, at $<$1~ms inference. Its training requires site-specific channels at all $N_d$ RIS positions, i.e., prior site surveys or extensive RT; PRIMO needs no training data.

\section{Ray-Tracing Validation}
\label{sec:results}

\subsection{Simulation Environment}

\begin{figure}[t]
    \centering
    \includegraphics[width=0.82\columnwidth]{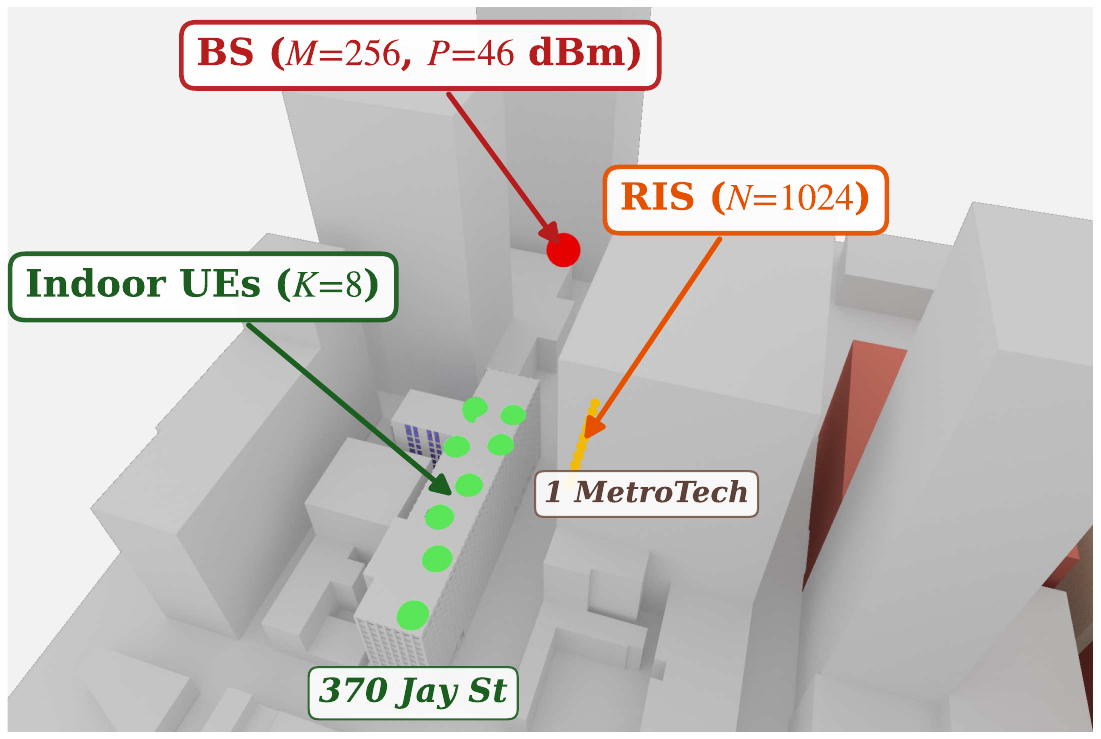}
    \caption{MetroTech Commons digital twin. BS (red) on rooftop, RIS candidates (orange) on the west facade of 1~MetroTech Center, indoor users (green) on the 9th floor of 370~Jay~St.}
    \label{fig:scenario}
        \vspace{-6pt}
\end{figure}

We evaluate the proposed framework using NYURay RT~\cite{kanhere2023_nyuray_icc, kanhere2025_nyuray} in a digital twin of MetroTech Commons with site-specific 3D building geometry and ITU-R P.2040 material assignments~\cite{ying2026_horama}. As shown in Fig.~\ref{fig:scenario}, the BS (3GPP TR~38.901 \S7.3 element pattern~\cite{3gpp_tr38901}) sits on a nearby rooftop, and a passive RIS on the west facade of 1~MetroTech Center, with outdoor LoS to the BS, redirects the BS signal toward UEs on the 9th floor of 370~Jay~St. Both BS and RIS are held at the same physical aperture across the two bands with $0.5\lambda$ spacing, so element counts scale as $f^2$: at FR1(C), $M\!=\!256$ ($16\!\times\!8$ dual-polarized VH) and $N\!=\!1024$ ($32\!\times\!32$); at FR3, $M\!=\!1600$ ($40\!\times\!20$) and $N\!=\!6400$ ($80\!\times\!80$), about $6.25\times$ more elements at both arrays. Table~\ref{tab:params} lists the remaining parameters.

\begin{table}[t]
\centering
\caption{Simulation parameters.}
\label{tab:params}
\setlength{\tabcolsep}{2pt}
\begin{tabular}{l|l|l|l}
\hline
\textbf{Parameter} & \textbf{Value} & \textbf{Parameter} & \textbf{Value} \\
\hline
Freq.\ $f_c$ & 6.75/16.95~GHz & BS power $P$ & 46~dBm~\cite{3gpp_tr38913} \\
BW / NF & 20~MHz / 7~dB & UEs $K$ & 4--20, height 44.5~m \\
BS ant.\ $M$ & 256/1600, $0.5\lambda$ & RIS $N$ & 1024/6400, $0.5\lambda$ \\
RT engine & NYURay~\cite{kanhere2025_nyuray} & Scene & MetroTech Commons \\
Target bldg. & 370 Jay St. & RIS track & $L=40$~m, $N_d\!=\!10$ \\
BS height & 38~m & RIS height & 45~m \\
\hline
\end{tabular}
\vspace{-15pt}
\end{table}

All channels are generated via NYURay~\cite{kanhere2025_nyuray}, originally calibrated against measurements at 28, 73, and 142~GHz~\cite{kanhere2023_nyuray_icc}; the ITU-R P.2040 concrete and glass parameters used here are independently validated against FR1(C)/FR3 penetration-loss measurements~\cite{rappaport2024_fr1c, shakya2025_umi_icc, ying2025_inf_icc}. The RT uses specular reflection, refraction, and diffraction (concrete 20~cm, glass 12~mm)~\cite{itu_p2040} at interaction depth~4 for $\vect{h}_{d,k}$ and $\vect{f}_k$, depth~3 for $\mat{G}$. Narrowband channel vectors follow from $H(f_c) = \sum_\ell a_\ell e^{-j2\pi f_c \tau_\ell}$; flat fading holds since $W/f_c < 0.3\%$. Perfect cascaded CSI is assumed throughout~\cite{wu2019_ris_bf, wu2021_ris_survey}, with estimation overhead discussed in Sec.~\ref{sec:conclusion}.

\begin{table}[t]
\centering
\caption{Minimum user rate (bps/Hz) versus indoor user count $K$ at FR1(C) and FR3.}
\label{tab:results}
\setlength{\tabcolsep}{4pt}
\begin{tabular}{c|cc|cc|c}
\hline
& \multicolumn{2}{c|}{FR1(C) (6.75~GHz)} & \multicolumn{2}{c|}{FR3 (16.95~GHz)} & \\
$K$ & AO-EW & \textbf{PRIMO} & AO-EW & \textbf{PRIMO} & FR3/FR1(C) \\
\hline
4  & 10.93 & \textbf{10.94} & 8.21 & \textbf{8.50} & $0.78\times$ \\
8  & 5.62  & \textbf{5.83}  & 5.86 & \textbf{7.23} & $1.24\times$ \\
12 & 1.03  & \textbf{1.38}  & 4.63 & \textbf{6.49} & $4.70\times$ \\
16 & 0.64  & \textbf{0.81}  & 1.19 & \textbf{2.26} & $2.79\times$ \\
20 & 0.37  & \textbf{0.41}  & 0.99 & \textbf{1.60} & $3.85\times$ \\
\hline
\end{tabular}
\begin{minipage}{\columnwidth}
\footnotesize\raggedright
\vspace{3pt}
Wall-clock on RTX~4090 at $K\!=\!8$: AO-EW 71\,/\,6133~s, PRIMO 4\,/\,57~s (FR1(C)/FR3); warm-start tracking $\approx\!2$~ms. Rightmost column: PRIMO FR3-over-FR1(C) ratio.
\end{minipage}
\end{table}

\begin{table}[t]
\centering
\caption{Minimum user rate (bps/Hz) for all methods at $K\!=\!8$.}
\label{tab:k8_methods}
\setlength{\tabcolsep}{4pt}
\begin{tabular}{l|cc|c}
\hline
\textbf{Method} & \textbf{FR1(C)} & \textbf{FR3} & \textbf{Time (FR1(C)/FR3)} \\
\hline
No RIS                                & 0.00          & 0.00          & --              \\
Random Phase                          & 0.34          & 0.36          & $<$0.1~s        \\
Manifold-RCG~\cite{yu2020_manifold_ris} & 1.01          & 0.47          & 0.3 / 23~s      \\
BCD-SDR~\cite{luo2010_sdr}            & 1.01          & 0.92          & 0.1 / 65~s      \\
WMMSE+AO~\cite{shi2011_wmmse}         & 1.03          & 0.73          & 3.2 / 5257~s    \\
AO-EW~\cite{wu2019_ris_bf}            & 5.62          & 5.86          & 71 / 6133~s     \\
GLNN~\cite{glnn_wcl2024} (in-dist.)   & 6.39          & 7.79          & train 935 / 5689~s, infer $<$1~ms \\
\hline
\textbf{PRIMO (proposed)}             & \textbf{5.83} & \textbf{7.23} & 4 / 57~s        \\
PRIMO-oracle$^{*}$                    & 6.22          & 7.62          & 13 / 253~s      \\
\hline
\end{tabular}
\begin{minipage}{\columnwidth}
\footnotesize\raggedright
\vspace{3pt}
$^{*}$PRIMO-oracle uses $B\!=\!2000$, $T\!=\!800$ ($\sim\!12\times$ default compute), the best point PRIMO can reach. GLNN slightly exceeds it in-distribution.\\
\end{minipage}
\vspace{-15pt}
\end{table}

\subsection{Algorithm Comparison}

Table~\ref{tab:results} reports the minimum user rate of PRIMO and AO-EW~\cite{wu2019_ris_bf}, the strongest training-free classical baseline, across $K\!\in\!\{4,8,12,16,20\}$; Table~\ref{tab:k8_methods} adds a method-by-method comparison at $K\!=\!8$. UEs are drawn in descending channel-strength order from a candidate pool on the 9th floor of 370~Jay~St.

At $K\!=\!8$, three groupings emerge in Table~\ref{tab:k8_methods}. (i)~BCD-SDR~\cite{luo2010_sdr}, WMMSE+AO~\cite{shi2011_wmmse}, and Manifold-RCG~\cite{yu2020_manifold_ris} collapse to $\le\!1.03$~bps/Hz: their phase updates do not converge well at $N\!=\!6400$ and their fairness-blind precoders amplify worst-user IUI. (ii)~True element-wise AO-EW with RZF precoding and 30 random restarts is the strongest training-free classical method ($5.62/5.86$~bps/Hz, FR1(C)/FR3). (iii)~PRIMO ($5.83/7.23$) outperforms AO-EW by 4\%/23\%, and PRIMO-oracle ($6.22/7.62$, $\sim\!12\times$ compute) marks the best feasible point gradient-based search can reach. GLNN~\cite{glnn_wcl2024} attains $6.39/7.79$~bps/Hz in-distribution on its own training scene, slightly above PRIMO-oracle, but must be retrained for each new scene, BS placement, aperture, or $K$.

Turning to the algorithmic gap with $K$ in Table~\ref{tab:results}, at $K\!=\!4$ both methods saturate the FR1(C) array (10.93 vs.\ 10.94~bps/Hz). The PRIMO/AO-EW rate ratio grows mildly with $K$ at FR1(C) (1.00--1.34) and more steeply at FR3 (1.04, 1.23, 1.40, 1.90, 1.62 across $K=4,\ldots,20$), peaking at $K\!=\!16$ where the joint $(\mat{W}, \vect{\theta})$ update at FR3's $N\!=\!6400$ has the most room to suppress IUI. The joint-update gain is thus a few percent to $\sim\!2\times$, not orders of magnitude; the residual gap to PRIMO-oracle ($0.39$~bps/Hz at $K\!=\!8$) reflects the cost of serving multiple UEs through a single shared $\vect{\theta}$ under finite-aperture IUI.

\begin{figure}[t]
    \centering
    \includegraphics[width=0.86\columnwidth]{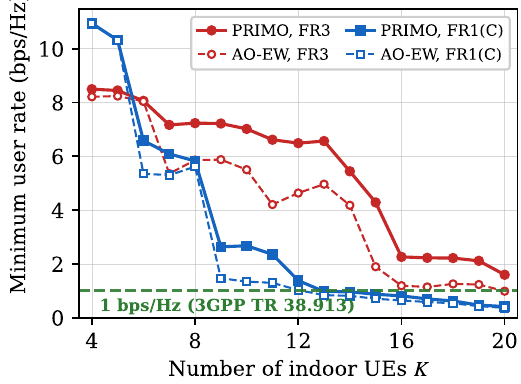}\vspace{-2mm}
    \caption{Minimum user rate vs.\ indoor user count $K$ for PRIMO and true AO-EW (markers at each integer $K$). Green dashed line: 1~bps/Hz edge-user threshold~\cite{3gpp_tr38913}. No-RIS is zero for $K\!\ge\!2$ and omitted.}
    \label{fig:k_curve}
    \vspace{-15pt}
\end{figure}

\subsection{Discussion}

Two observations emerge from the RT results.

(1) The FR3-versus-FR1(C) advantage is load-dependent, not constant. At $K\!=\!4$ FR1(C) leads by $1.29\times$ (Table~\ref{tab:results}) because the 7.99~dB free-space path-loss penalty at FR3 hits before the larger RIS has enough simultaneous users to exploit its extra spatial DoF; from $K\!=\!8$ on FR3 leads, peaking at $4.70\times$ ($K\!=\!12$), a multi-user capacity-scaling benefit rather than a constant single-link gain.

(2) RIS placement is a first-order design variable specific to O2I: moving the RIS along the facade changes which building material (concrete vs.\ glass) the RIS-to-UE path traverses, producing channel-quality jumps that no phase optimization can recover. Across the $N_d\!=\!10$ candidates, adjacent-facade placement with unobstructed outdoor LoS to the BS yields the lowest cascaded path loss at both bands, and the penalty for a poor position grows at FR3 where the narrower beam increases position sensitivity.

A 1~bps/Hz minimum user rate is the edge-user fairness floor invoked in 3GPP TR~38.913~\cite{3gpp_tr38913}, below which a UE cannot sustain basic services. Defining the service capacity $K_{\mathrm{serv}}$ as the largest user count whose worst UE clears this floor, both AO-EW and PRIMO reach $K_{\mathrm{serv}}\!=\!12$ at FR1(C), while at FR3 AO-EW reaches $K_{\mathrm{serv}}\!=\!19$ and PRIMO $K_{\mathrm{serv}}\!=\!23$, a $1.92\times$ FR3-over-FR1(C) capacity lift attributable to the $6.25\times$ element budget at fixed physical aperture.

\section{Conclusion}
\label{sec:conclusion}

Calibrated ray tracing of a MetroTech Commons digital twin shows that, at constant BS and RIS physical aperture, the FR3-over-FR1(C) advantage is load-dependent: FR1(C) leads at light load, while at $K\!=\!12$ FR3 delivers a 4.70-fold minimum user rate as its roughly six-fold larger element count suppresses IUI. PRIMO improves the minimum user rate by up to 90\% over a fully converged AO baseline while running about 100 times faster on a single GPU, and supports 92\% more indoor users at FR3 above the 1~bps/Hz service floor, with deployment geometry a first-order design variable alongside phase optimization. Future work includes practical cascaded-CSI estimation~\cite{zheng2022_ris_ce_survey}, cross-scene validation, user scheduling at high load, and multi-RIS or sub-array partitioning to relax the single-shared-$\vect{\theta}$ bottleneck.

\section*{Acknowledgment}
This work was supported by the NYU WIRELESS Industrial Affiliates Program, NYU Technology Opportunities \& Ventures, the NYU Tandon ECE Ph.D. Fellowship, and NSF Grants 2216332, 2234123, and 2403287.

\bibliographystyle{IEEEtran}
\bibliography{references}

\end{document}